# On the impact of Masking and Blocking Hypotheses for measuring efficacy of new tuberculosis vaccines

Sergio Arregui *[1,2], Joaquín Sanz[1,3,4], Dessislava Marinova[5,6], Carlos Martín[5,6,7], Yamir Moreno[1,2,8]

1 Institute for Biocomputation and Physics of Complex Systems (BIFI), University of Zaragoza, Spain
2 Department of Theoretical Physics, University of Zaragoza, Spain
3 Sainte-Justine Hospital Research Centre, Montreal, Canada.
4 Department of Pediatrics, University of Montreal, Canada.
5 Department of Microbiology, Faculty of Medicine, University of Zaragoza, Spain
6 CIBER Enfermedades Respiratorias, Instituto de Salud Carlos III, Madrid, Spain.
7 Service of Microbiology, Miguel Servet Hospital. IIS Aragon.
8 Complex Networks and Systems Lagrange Lab, Institute for Scientific Interchange, Turin, Italy

## ABSTRACT

Over the past 60 years, the *Mycobacterium bovis* bacille Calmette-Guérin (BCG) has been used worldwide to prevent tuberculosis (TB). However, BCG has shown a very variable efficacy in different trials, showing a wide range of protection in adults against pulmonary TB. Previous studies indicate that this failure is related to pre-existing immune response to antigens that are common to environmental sources of mycobacterial antigens and *Mycobacterium tuberculosis*. Specifically, two different mechanisms have been hypothesized: the masking, (previous sensitization confers some level of protection against TB), and the blocking (previous immune response prevent vaccine taking of a new TB vaccine), effects. In this work we introduce a series of models to discriminate between masking and blocking mechanisms and address their relative likelihood. The application of our models to interpret the results coming from the BCG-REVAC clinical trials, specifically designed for the study of sources of efficacy variability yields estimates that are consistent with high levels of blocking (41% in Manaus -95% C.I. [14%-68%]- and 96% in Salvador -95% C.I. [52%-100%]-), and no support for masking to play any relevant role in modifying vaccine's efficacy either alone or aside blocking. The quantification of these effects around a plausible model constitutes a relevant step towards impact evaluation of novel anti-tuberculosis vaccines, which are susceptible of being affected by similar effects if applied on individuals previously exposed to mycobacterial antigens.

## INTRODUCTION

Despite all the efforts in the fight against TB accomplished during the last decades, the disease still remains a major cause of morbidity and mortality worldwide, being responsible for a million and a half deaths per year all around the world [1]. The increasing emergence of multi drug and extremely drug resistant strains [2] or the association between TB and VIH [3,4] constitute serious epidemiological threats that evidence the necessity of further public health measures and pharmacological resources against the disease.

Among all the possible epidemiological interventions that could contribute to the desired goal of TB eradication, the introduction of a novel preventive vaccine is currently thought to be able to offer the highest and most immediate impact on disease burden reduction, since the efficacy of the current TB vaccine BCG is reduced, and only consistent in protecting infants, specially from the most severe forms of meningeal and miliary TB [5]. Accordingly, nowadays there exist more than fifteen different research teams worldwide developing as many novel experimental vaccine candidates designed to substitute BCG, or, at least, to be applied as a booster in BCG vaccinated individuals in order to enhance the protective effects of the old vaccine [6].

BCG fails to provide consistent protection to the pulmonary forms of the disease, especially in adults [7], who are the main contributors of overall disease spreading. Consequently, an accurate evaluation of the BCG impact under different conditions – population susceptibility, geography, environmental exposure, etc- is essential: firstly, it will allow to assess the efficiency of BCG as a reference vaccine and secondly, it will provide new guidelines and methodological tools to better evaluate the potential efficacy of the newly developed TB vaccines. The highly variable and apparently inconsistent results obtained in BCG's efficacy tests and meta-analysis have been subject of intense scientific controversy [5,8]; and the use of BCG during the 20th century has been largely argued [9,10].

The hypothesized causes underlying the observed variability of BCG efficacy in different settings include differences between the BCG strains [11], genetic, epi-genetic or socio-economical differences between populations, study quality, parasitic co-infections, etc [6]. In addition, multi-variate meta-analysis of BCG efficacy determination studies consistently determine that latitude is a variable showing a direct correlation with BCG performance [10,12,13,14], pointing to the existence of latitude-driven mechanisms influencing it, rather than other possible explanations related, for example, to the ethnicity of the tested populations [15]. Among these possible mechanisms, the hypothesis that agglutinates a greater consensus points to the existence of a complex, latitude-dependent immunological process of environmental sensitization (ES) to mycobacterial antigens which might interfere with the observed action of BCG vaccine in different ways. The hypothesis of ES being the source of BCG efficacy variability has been backed up by different epidemiological observations [5,16,17,18].

ES is thought to have its origin in the exposure of individuals either to non tuberculous mycobacteriae (NTM) -whose antigenic similarity to MTB [19,20] is able to cause cross reactivity in the human immune system [21,22]-, or to the reservoir of latent infection of MTB itself (and other closely related bacteria within the MTB-complex). Even if the diversity among the different putative sources of ES (both in terms of prevalence depending on latitude and in terms of cross reactivity levels) portrays a complex landscape that makes specially ventured to attribute the geographical patterns of BCG efficacy variation to a single factor (as for example, to an hypothetical increase in NTM prevalence levels next to the equator [21,22,23], which has been demonstrated to be inaccurate for some species [24]), it seems clear that overall levels of ES increase both with closeness to equator and subjects' age at the time of vaccination.

Two different mechanisms have been theorized on how this exposition to mycobacteriae would affect the response of the host to a vaccine like BCG [25]. The masking hypothesis postulates that environmental sensitization confers a significant protection against TB in such a way that a vaccine can barely offer an additional level of protection [25,26]. Another possible effect is that prior ES may trigger an immune response capable of blocking the assimilation of the vaccine by the host, either if it's a live-attenuated vaccine or if it's a booster. This is known as the blocking hypothesis [25,27]. These two effects have the potential to explain, to a large extent, the variability observed in the trials performed, that is, both the dependence of BCG efficacy on age at the time of vaccination – as an individual gets older its exposition to mycobacteriae increases – and its geographical variations. Relevantly enough, if BCG were affected by masking, the temporal patterns of observed protection loss after vaccination of newborns would be different in different areas subject to different levels of ES. That would not be the case for blocking, for which it would be enough to apply the vaccine right after birth, when ES has not taken place, no matter the geographical location.

In order to elucidate which of these mechanisms drives the observed variability between different studies, and up to what extent, BCG-REVAC trials were designed to discriminate these two effects on BCG performance when applied on individuals of dissimilar ages in different cities of different latitudes within the same country (Salvador and Manaus, in Brazil) [8,28,29]. After the analysis of BCG-REVAC trials, Barreto et al. observed that the efficacy of the vaccine, when applied to newborns and measured later in life did not show a strong geographic variation, at variance to what it was observed when BCG was applied at school age [8]. In addition, for BCG revaccination at school age, protection was modest in Salvador and absent in Manaus [8]. These observations suggested that blocking plays a more relevant role than masking on explaining BCG efficacy variations. However, even if the design of BCG-REVAC trials allowed to qualitatively asses the greater relevance of the blocking effect as compared to masking, no actual quantification of these two effects and their relative role has been provided up to now. In this sense, after the work by the BCG-REVAC consortium, several questions remain unanswered, as we do not know 1) what's the relative likelihood of both hypothetical mechanisms when trying to explain the observed results of the trial, 2) how much predictive power would a full model containing both effects gain with respect to single effects scenarios (masking or blocking alone) 3) whether significantly different combinations of masking and blocking strengths could be similarly compatible with the observations derived from the trials or not, and, very relevantly, 4) what are the intensities of blocking and masking effects, and their confidence intervals, yielding a most significant agreement with the data.

In this paper, we introduce a family of mathematical models to interpret the results from BCG-REVAC under the light of masking and blocking effects, in order to contribute to answer the aforementioned questions within the limitations imposed to the reduced statistical power derived from the reduced number of trials studied. By confronting our model against the results of the BCG-revac studies, we are able to measure up to which extent these effects are sufficient to explain the efficacies measured [8], as well as to quantify the specific masking, blocking and immunity waning effects yielding best fit estimates for the efficacies measured, comparing to this end the likelihoods of three different modeling scenarios: a first model in which both effects take place concurrently, a second model only considering blocking and a third one containing only masking. It is worth noticing that translating the trials results into quantitative estimations of blocking and masking strengths is a relevant step that enables mathematical models of disease spreading aimed at evaluating vaccine impacts to take into account these effects, both in what regards the dependence of BCG efficacy itself on individuals' age and geographical areas and also in what concerns the plausible ranges of blocking and masking that any novel vaccine might eventually suffer.

**METHODS**

**Data analyzed: the Brazilian BCG-REVAC clinical trials**

BCG-REVAC consisted of a set of cluster-randomized trials involving more than 200,000 school-aged children in the Brazilian cities of Manaus and Salvador, whose principal aim was evaluating the effectiveness of BCG under different vaccination protocols.

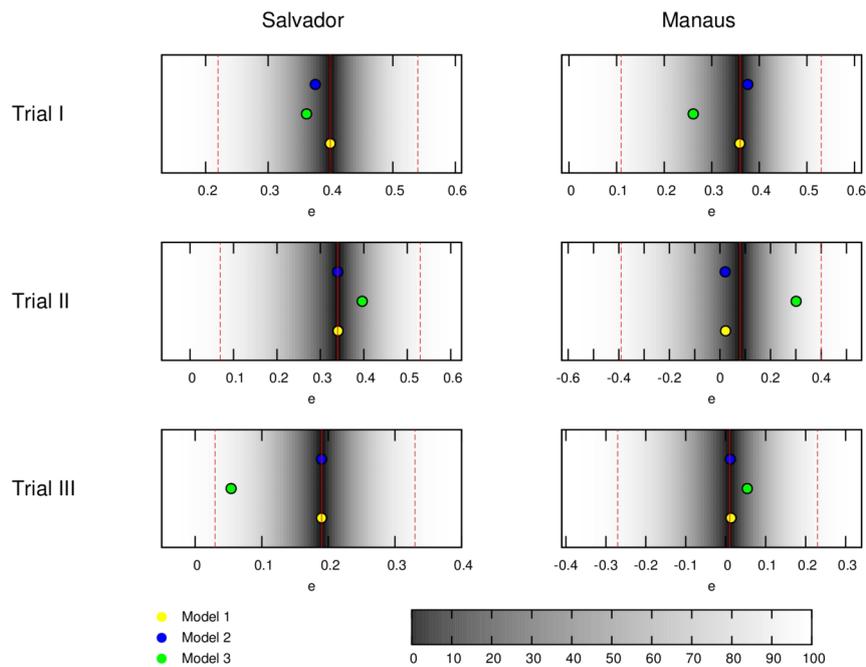

Figure 1: Best fit estimates for each trial by models 1, 2 and 3 (yellow, blue and green dots, respectively) for the trials conducted in the BCG-revac study. The colormap represents the probability of obtaining a more extreme value of the efficacy, according to the distributions considered. The probability of zero marks the central estimate (red, continuous line) while the dashed red lines mark the 95% C.I. reported by [6].

The enrolled population of the study consisted of non-infected school children between 7 and 14 years old at the moment of randomization. Within this population, individuals presenting a positive BCG scar are separated from the rest, distinguishing, this way, the enrolled individuals who were vaccinated at birth from those who were not. Each group is then split into an intervention and a control group; individuals in the intervention group were vaccinated within the context of the trial. Summing up, there are 4 cohorts in each city: non vaccinated (1), vaccinated after birth (2), firstly vaccinated at school age during the trial (3), and revaccinated, after a first dose applied after birth, in the trial too (4). Upon such classification of enrolled individuals in cohorts, the effectiveness of BCG vaccination strategies was measured by comparing the TB incidence rate within an end-point associated to active disease in the four cohorts, according to three different types of trials: Trial I: BCG at birth vs. no intervention (cohort 2 vs. 1). Trial II: BCG first dose at school age vs. no intervention (cohort 3 vs. 1). Trial III: revaccination at school age vs. first dose at birth only (cohort 4 vs. 2).

**A model to describe BCG efficacy variation: masking, blocking and immunity waning.**

The six clinical trials conducted within the framework of BCG-REVAC study –three types of trials per two cities- output efficacies that span from 1% to 40% protection (see figure 1, red continuous lines). In order to explain this variability we propose a model according to which the different protection levels found in each of the four cohorts in the study, schematically shown in figure 2, result from the interplay between the intrinsic vaccine efficacy, its temporal waning patterns, masking and blocking effects. These three mechanisms of vaccine protection shifts are ultimately responsible for vaccine's performance variation, either in space (i.e. between the two cities of the study) or in time, as function of individuals' age at the moment of vaccination and/or the time passed between vaccination and observation.

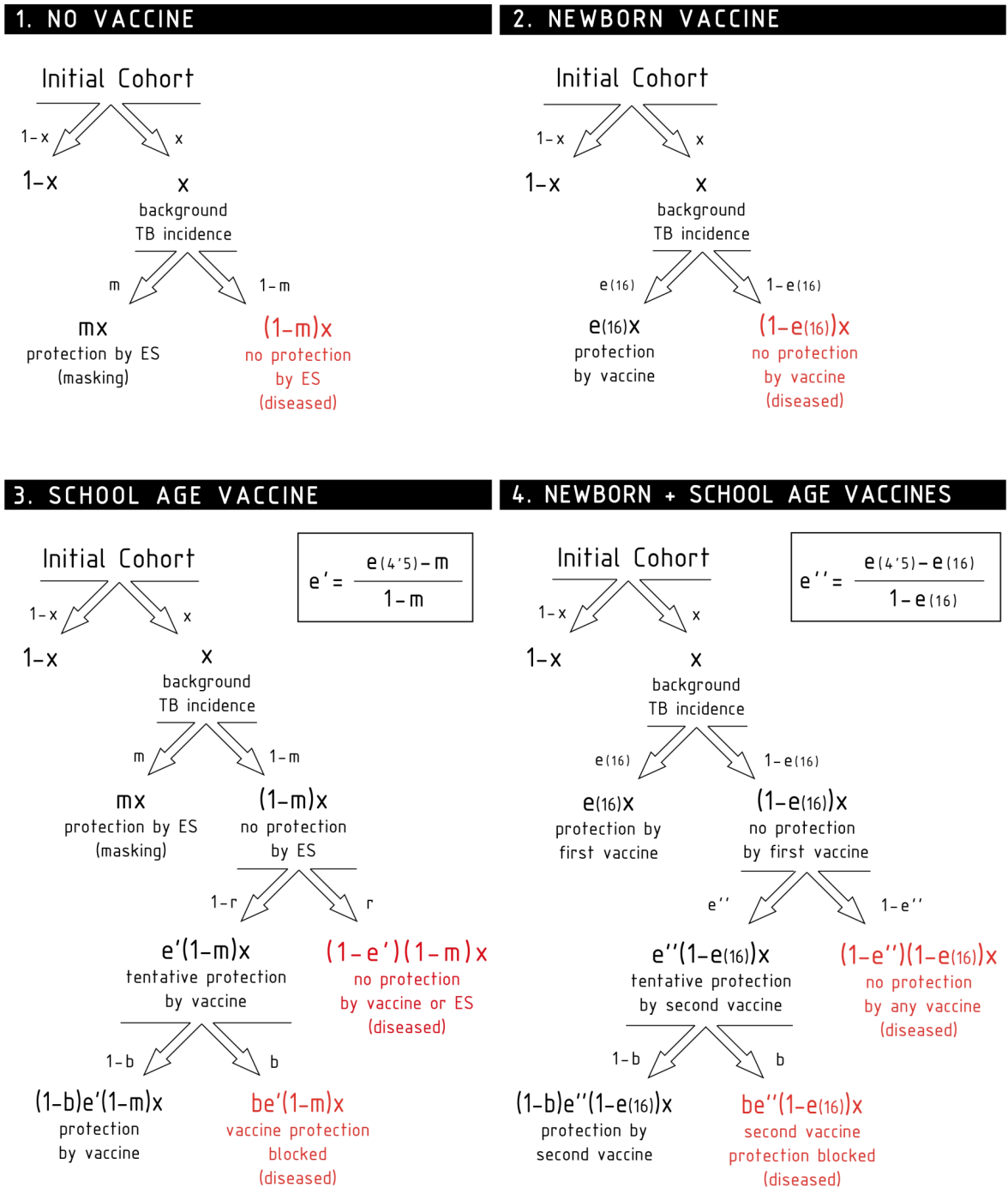

**Figure 2: Scheme of the different contributions to the disease risk for each cohort.**

First of all, in absence of masking or blocking, a naive vaccinated individual will receive a protection level, right after vaccination, that we call $e(0)$. As time after vaccination goes by, this protection level will wane up to $e(t) < e(0)$, generally speaking. This implies that, if we deal with a population in which the incidence rate of new TB cases per unit time is equal to $x$; $t$ years after vaccination this rate is modified to $(1-e(t))x$, provided that no additional effects take place. Taking that into account, a protective vaccine will have positive efficacy values $e(t) \in (0,1]$, being also possible for a (failed) vaccine to have a negative efficacy if it augments the disease risk among vaccinated individuals instead of reducing it. In our model, the time waning patterns of the intrinsic vaccine efficacy do not depend on the geographical area, but just on time since vaccination, which approximately is, in average, 4.5 years for school age vaccination (cohort 3) and 16 years for newborn vaccination (cohort 2), which implies the consideration of two intrinsic efficacy parameters: $e(4.5)$ and $e(16)$.

Besides vaccination, ES can also support protection against disease through the masking mechanism. The masking level, denoted by $m$, is a protection parameter formally equivalent to the intrinsic vaccine efficacy (thus verifying $m \in (0,1]$ for a protective effect, and negative otherwise), whose effects are suffered by initially naive, non-vaccinated individuals subject to ES. Thus, in principle, the longer the time an individual has been exposed to ES –i.e. the older the individual is at the moment of observation-, the higher is the masking-related protection she might show. Masking is also a geography-dependent effect, which forces us to consider two masking parameters: $m^M$ for Manaus and $m^S$ for Salvador. The dependence of these parameters on age cannot be resolved, since all the cohorts analyzed in the study has approximately the same age.

Additionally, if $e(t)$ describes the protection provided by the vaccine to a naive individual in absence of masking or blocking $t$ years after vaccination, we also need to describe how this protection is modified if the vaccine is applied to non-naive subjects. If an individual's immune system has been stimulated prior to vaccination (either by masking like in cohort 3, or by a previous vaccine, like in cohort 4 before the second dose), and consequently she is partially protected against the disease, it is unrealistic to assume that the full effect of the new dose is additive [25]. Instead of that, our model considers that a vaccine dose applied on a previously protected individual will contribute up to resetting of the initial protection levels $e(0)$, provided that no blocking takes place. This implies that, right after the school age vaccination, in cohort 3, if the vaccine is not blocked ($b=0$, see below), it will have a protective effect $e'$ that will be concurrent with the masking protection $m$ so as to reduce the disease risk to $[1-e'][1-m]x$. Our estimation of $e'$ comes from assuming that such disease risk must equate what we would observe if a vaccine of full efficacy were applied on naive individuals, and observed 4.5 years later:

$$[1-e'][1-m]x = [1-e(4.5)]x \rightarrow e' = \frac{e(4.5)-m}{1-m} \qquad (1)$$

Similarly, the school-age dose at cohort 4, will add to the protection provided by the newborn dose $e(16)$, an additional factor $e''$, To obtain $e''$, we assume that, if not blocked, the disease risk achieved by both vaccines together ($[1-e(16)][1-e'']x$ is equivalent to the disease risk reached by the same vaccine, if applied on unprotected invduals, 4.5 years after vaccination:

$$[1-e(16)][1-e'']x = [1-e(4.5)]x \rightarrow e'' = \frac{e(4.5)-e(16)}{1-e(16)} \qquad (2)$$

Finally, vaccine intrinsic efficacy can be blocked by prior ES; an effect that we model through the blocking probability $b \in [0,1]$, where $b=0$ means that no blocking appears, while $b=1$ stands for a totally blocked vaccine, meaning that vaccinated individuals will only posses the protection level that they already had before vaccination. Blocking is also a geography-dependent factor, since it is considered a consequence of ES as well, which forces us to distinguish $b^M$ and $b^S$ for Manaus and Salvador, respectively. Unlike masking, blocking does not depend on the age of the individuals at the moment of observation, but on their age at the moment of vaccination. In this case we study cohorts vaccinated at two moments in life –at birth and at the beginning of the trials- being the first of these cases (the newborn vaccination) considered blocking-free, as it is assumed that when the vaccine is applied immediately after birth, there is no place for prior ES.

Taking all these effects into account, we are left with a set of six independent parameters

$\vec{P} = \{e(4.5), e(16), m^M, b^M, m^S, b^S\}$ to describe the variability observed in the trials, either temporal or geographical, under the light of blocking and masking effects, concurrently. The temporal trends of the level of protection of each cohort are schematically shown in figure 3.

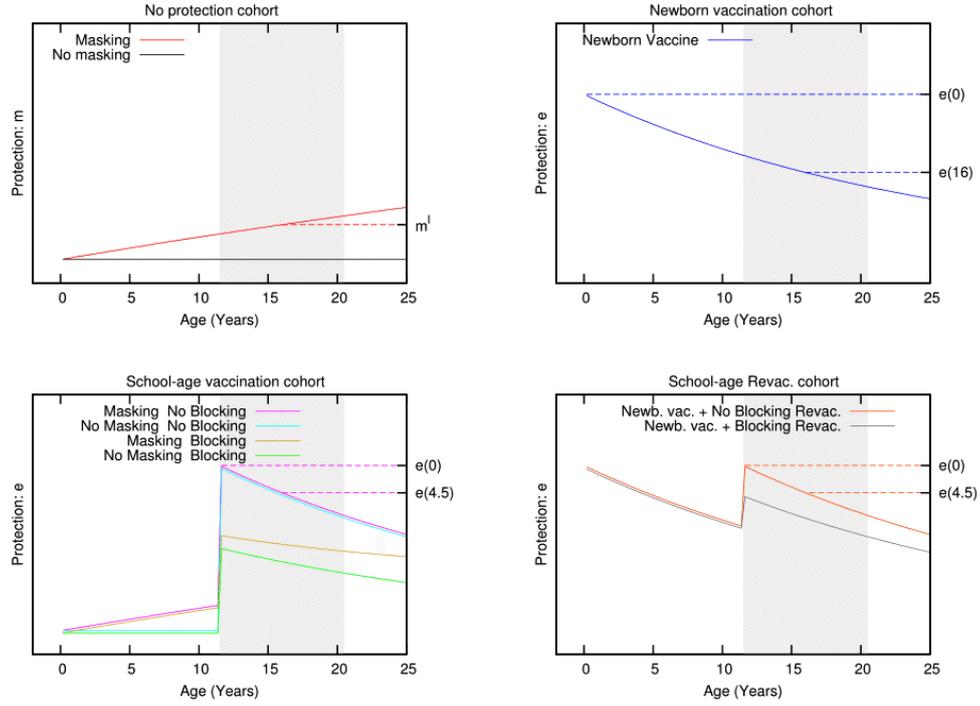

**Figure 3: Scheme for the temporal evolution of the level of protection for the 4 cohorts considered in the work, according to the different vaccination and ES possibilities. Cohort one: protection against disease of non-vaccinated individuals can only be due to masking, which is an increasing function with age. Cohort 2: Newborn vaccination cohort: individuals are vaccinated right after birth, which provides a protection that overcomes any possible masking effect, cannot be blocked by ES and wanes with time. Cohort 3: school age, first dose: masking, if present, can initially protect individuals. Then, at the moment of vaccination, if not blocked, the vaccine will overcome masking protection up to the initial value $e(0)$, which then will wane. Cohort 4: School age re-vaccination cohort: after a newborn, unblocked vaccination, a second dose is applied, which, if not blocked, will reset the initial protection levels provided by the vaccine. The grey shaded area represents the age window of the individuals enrolled in the study.**

In the following, we will refer to this full model as model 1. In figure 2, we represent the variations on the disease rates provoked by each effect that takes place in each cohort according to model 1. Summing all the possible *contributions* to the development of active disease for each cohort, we derive the general disease rates characterizing each cohort of one city as follows:

$$d_1^l = (1 - m^l)x$$
$$d_2^l = [1 - e(16)]x$$
$$d_3^l = (1 - b^l)[1 - e(4.5)]x + b^l(1 - m^l)x \quad (3)$$
$$d_4^l = (1 - b^l)[1 - e(4.5)]x + b^l(1 - e(16))x$$

where the superscript indicates location, and $x$ the incidence rate observed in the population.

From (3), it is immediate to derive the expressions for the observed efficacies $\bar{e}$ of each trial according to model 1, which read as:

$$M_1 : \begin{cases} \bar{e}_I^l = 1 - d_2^l/d_1^l = \dfrac{e(16) - m^l}{1 - m^l} \\ \bar{e}_{II}^l = 1 - d_3^l/d_1^l = \dfrac{e(4.5) - m^l}{1 - m^l}(1 - b^l) \\ \bar{e}_{III}^l = 1 - d_4^l/d_2^l = \dfrac{e(4.5) - e(16)}{1 - e(16)}(1 - b^l) \end{cases} \qquad (4)$$

The system of equations (4) represents a full model for the vaccine efficacies observed during BCG-REVAC trials, which is based on the assumption that the sources of geographical variability for BCG's performance are both masking and blocking effects. From the full model, two reduced versions can be conceived: a masking-free model (model 2 in the following) in which $m^M = m^S = 0$, and a blocking free model in which $b^M = b^S = 0$ (model 3). The efficacies associated to each trial, for models 2 and 3 straightforwardly read as follows:

$$M_2 : \begin{cases} \bar{e}_I^l = e(16) \\ \bar{e}_{II}^l = e(4.5)(1 - b^l) \\ \bar{e}_{III}^l = \dfrac{e(4.5) - e(16)}{1 - e(16)}(1 - b^l) \end{cases} \qquad (5)$$

$$M_3 : \begin{cases} \bar{e}_I^l = \dfrac{e(16) - m^l}{1 - m^l} \\ \bar{e}_{II}^l = \dfrac{e(4.5) - m^l}{1 - m^l} \\ \bar{e}_{III}^l = \dfrac{e(4.5) - e(16)}{1 - e(16)} \end{cases} \qquad (6)$$

By considering these three models, our approach allows quantifying and comparing the plausibility of blocking and masking hypotheses to potentially explain the variation in BCG efficacy trials observed in the controlled setup conceived in the BCG-REVAC trials, taking into account the non-linearities associated to each mechanism, which play a central role in the derivation of Equations (4-6).

**Models solution: parameters estimation and confidence intervals.**

In order to identify the set or sets of parameters yielding a best fit for the efficacies observed in BCG-REVAC trials, we compare the model prediction associated to any parameter set $\vec{P}$ to a set of empirical probability distributions derived from BCG-REVAC data. From each of the confidence intervals reported in [6] we build a two-piece normal distribution [30] for each trial reported, centered in the reported values $\left[\bar{e}_i^l\right]_{BCG-REVAC}$ (for location $l \in \{Manaus, Salvador\}$ and trial $i \in \{I, II, III\}$), and with asymmetric variances $\left[\sigma_i^l\right]^{\pm}_{BCG-REVAC}$ equal to one half the radius of the confidence intervals reported in [6], so, preserving the confidence levels of the intervals reported (see figure 1).

Once the empirical distributions have been defined, for each possible set of parameters and for each of the six trials we define the Z-score associated to the model prediction as:

$$Z_i^l(\vec{P}) = \dfrac{\left|\left[\bar{e}_i^l(\vec{P})\right]_{mod} - \left[\bar{e}_i^l\right]_{BCG-REVAC}\right|}{\left[\sigma_i^l\right]^{\pm}_{BCG-REVAC}} \qquad (7)$$

where $\left[\sigma_i^l\right]^{\pm}_{BCG-REVAC}$ will take each of its two possible values depending on the sign of $\left( \left[ \bar{e}_i^l(\vec{P}) \right]_{mod} - \left[ \bar{e}_i^l \right]_{BCG-REVAC} \right)$. From $Z_i^l(\vec{P})$, we define the corresponding p-values $p_i^l[Z_i^l(\vec{P})]$ as the probability of the empirical distributions reproducing BCG-REVAC data to have a Z-score $\tilde{Z}$ so that $|\tilde{Z}| > |Z_i^l(\vec{P})|$. This allows us to define the following likelihood function:

$$L(\vec{P}) = \prod_{l,i} p_i^l[Z_i^l(\vec{P})] \qquad (8)$$

to maximize so as to identify the model's parameters $\vec{P}^*$ more likely to yield the BCG-REVAC results. The global landscape of $L(\vec{P})$ is explored using a hill-climbing algorithm designed to identify all possible local maxima in the space of parameters. Finally, a Levemberg-Marquardt algorithm is used to find a more accurate value of the global maximum, if the latter is unique.

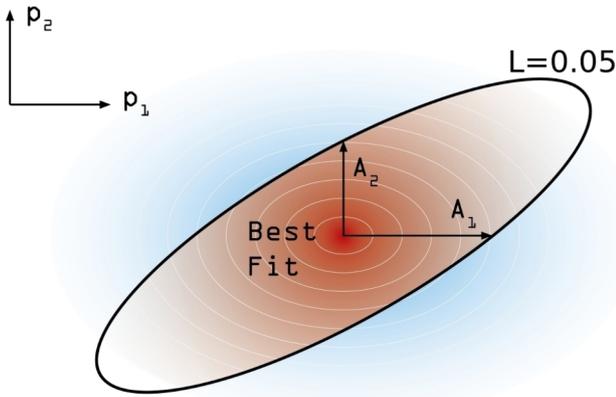

Figure 4: Confidence intervals estimation scheme. Degraded shades represent the joint probability density associated to the estimation of confidence intervals around the model best fit. The modulation coefficient $c$ is determined so as to make the brown area within the black line of $L(P) = 0.05$ to precisely accumulate the 95% of the total joint probability distribution.

In order to estimate the confidence interval associated to our model estimation, the following numerical procedure is performed. First, and starting from the maximum likelihood estimate $\vec{P}^*$, we move on each parameters' axis until a value of $L = 0.05$ is reached in each case. We call this increment $A_j$ ($j \in (1,6)$) (see figure 4). These values are not symmetrical, again, and so we distinguish between $A_j^+$ and $A_j^-$. Using these asymmetric widths, we construct a two-piece normal distribution for every parameter [30], centred in $P_o$ and having an asymmetric variance given by $\sigma_j^{\pm} = cA_j^{\pm}$, where $c$ is a common modulation coefficient. Besides, the distribution is truncated at 1. Finally we numerically estimate $c$ by generating sets of points in the parameter space whose coordinates in each axis are obtained from the split normal distributions mentioned for an initial guess of $c$. Through an iterative process we search the value $c = c^*$ for which a 95% of the points generated in the parameters space, yield efficacy estimations verifying $L(P) > 0.05$. Once we have found the optimal value of the scaling coefficient, the reported uncertainty of the j-th parameter corresponds to 95% C.I. given the distributions we have used.

**RESULTS**

In order to find the set or sets of parameters yielding best estimates of BCG efficacies according to our models we have performed a series of numerical optimization procedures seeking for likelihood maximization. First, we are interested in addressing whether a unique likelihood maximum exists across the parameter space of each model or whether, on the contrary, there exist multiple minima associated to different, yet comparable values of $L(\vec{P})$. This is an important question to address, as the existence of different maxima in a model would be eventually associated to the existence of different parameters sets describing vaccine's properties all of them compatible with the general model formulation. To solve that question, we performed an iterative hill-climbing algorithm starting from 20,000 random points across the parameter space for each model. As it can be seen in figure 5, while model 2 presents a unique likelihood maximum ($L(\vec{P}^*) = 0.53$), models 1 and 3, which contemplates masking, fails at providing a univocal vaccine's description associated to a unique solution from likelihood optimization.

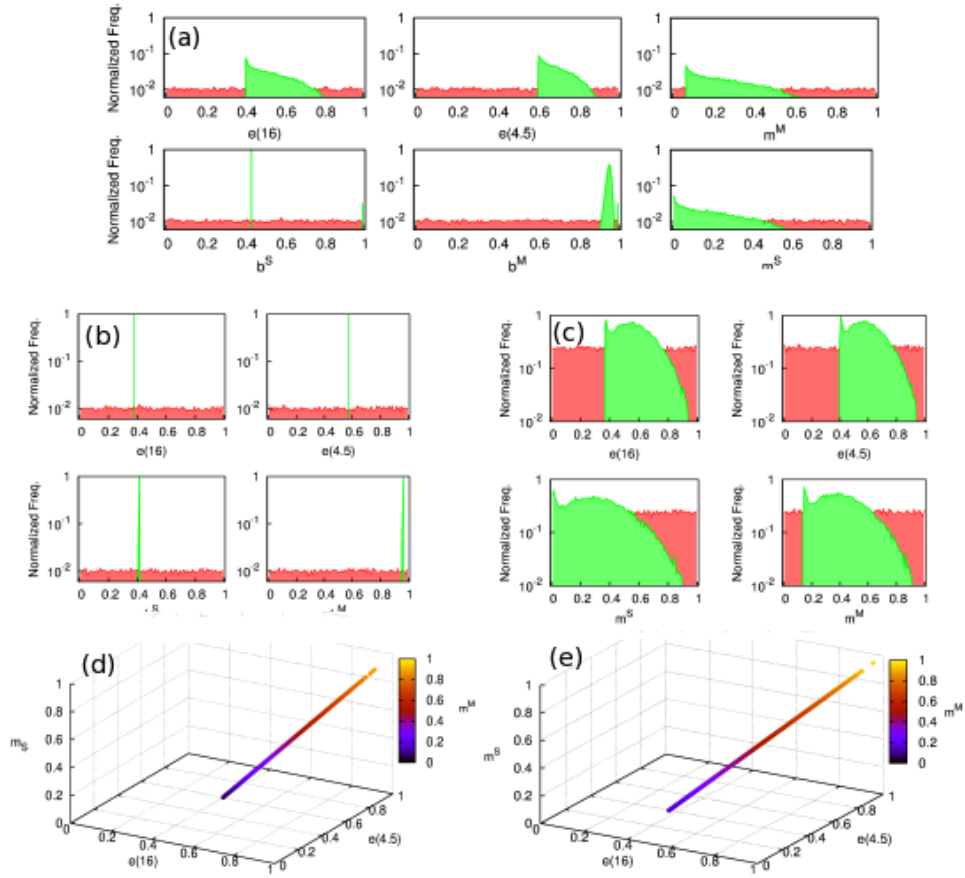

Figure 5: Panels a,b,c: Hill climbing algorithm distributions for models 1, 2 and 3, respectively. Starting from a series of randomly distributed points in the parameter space (their coordinates distributions are represented in red), a random displacement following a uniform distribution in the parameter space within a hyper-cube of size $\delta = 0.001$ is attempted at each time step, and accepted only if it corresponds to an increasing of the likelihood function $L(\vec{P})$. The algorithm stops when no further move is accepted after $N = 10^7$ rejected displacements (i.e. the function $L(\vec{P})$ reaches a maximum). In green, we see the peaked distribution of the end points of the algorithm around the solution of the models. Panels d and e: parameters cliff yielding quasi-constant values of maximum likelihood $L(\vec{P}^*) = 0.79$ for model 1 and $L(\vec{P}^*) = 0.002$ for model 3. As it can be seen in panel a and c, the model versions that contemplate masking are unable to provide a clear univocal vaccine description yielding maximum likelihood. The reason for this behavior is the existence of a region in the parameters space, represented in panel d and e, within which, likelihood is almost constant and close to its absolute maximum.

Instead of that, as we can see in figure 5, panels a,c,d and e models 1 and 3 present a *parameters cliff* across which model's likelihood is near to its maximum, and largely comparable ($L(\vec{P}^*) = 0.79$ for model 1, and $L(\vec{P}^*) = 0.002$ for model 2). Furthermore, a relative likelihood test comparing models 2 and 3 (that is, comparing blocking vs masking as exclusive mechanisms) yields a relative likelihood $L_3(\vec{P}^*)/L_2(\vec{P}^*) = 0.002/0.53 = 3.8 \cdot 10^{-3}$, which, considering that both models share the same amount of parameters, highlights again the inability of masking to provide a picture for vaccine efficacy variation as accurate as blocking does, as we can also see in figure 1, where the best fit provided by each model is presented as well.

If the analysis of model 3 and its comparison against model 2 allows us to discard masking as an autonomous mechanism able to explain the vaccine efficacy measured in the trials, it remains to be elucidated whether its consideration aside blocking in model 1 might still be able to significantly improve the fitting of the observed data. To answer this question, we conduct a simple likelihood ratio test in which the null and full models are, respectively models 2 and 1. From such test, we get that the statistic: $\chi^2 = -2\ln(L_2(P^*)/L_1(P^*)) = 0.80$, is a chi-square distributed variable with 2 degrees of freedom (difference between number of parameters of models 1 and 2) under the null, containing blocking alone, model. The obtained value does not allow to discard it even with a 50% confidence ($X^2(p=0.5, df=2)=1.39$), which indicates that masking is not just unable to provide an acceptable description of the observed data by itself but also makes no significant contribution to explain the variations in vaccine efficacies observed in the trials under study, when considered in addition to blocking. This is also reflected in the close estimates that are found for blocking parameters in models 1 and 2 (see table 1 and figure 5). Besides, if we analyze the combination of parameters that formed the cliff of maximum likelihood in model 1, we see that it consists in very similar levels of masking for the two different cities, which enters into conflict with the mentioned correlation between ES effects and closeness to equator.

| Parameter | Model 2 (only blocking) |
|---|---|
| $e(4.5)$ | 57.7% (46.8% to 68.6%) |
| $e(16)$ | 37.6% (29.3% to 45.8%) |
| $b^M$ | 96.4% (51.9% to 99.8%) |
| $b^S$ | 41.1% (14.2% to 68.0%) |

**Table 1: Optimal parameters of model 2. Models 1 and 3 are unable to provide a unique parameter set yielding maximum likelihood.**

In summary, our results point to blocking as the only plausible source of vaccine efficacy variation between the two mechanisms considered, validating the qualitative interpretation of the BCG-revac outcomes by Barreto et al. [8]. The best fit of model 2 yields a likelihood $L(\vec{P}^*)=0.53$, which corresponds to moderate blocking levels in Salvador ($b^S=0.41$ c.i. [0.14,0.68]) and to almost total blocking in Manaus ($b^S=0.96$ c.i. [0.52,1.00]), results that are, in turn, consistent with the assumed correlation between ES action strength and closeness to equator.

**DISCUSSION**

Understanding the mechanisms driving ES effects on BCG performance is a crucial task in the agenda towards the development of new tuberculosis vaccines. In this work, we have proposed a mathematical model that allows the quantitative evaluation of these two effects based on the BCG-REVAC trials performed in Brazil [8]. We have seen that the divergence in the measured efficacies of the trials is explained with high values of blocking, which concur with the qualitative discussion made in Barreto, M.L. et al. [8]. Furthermore, we have also observed for the first time that no alternative behavior of BCG is compatible with the observed data within the context of a model in which BCG's variability is entirely attributed to ES sensitization.

Admittedly, the range of applications of the results here exposed must be restricted to the provision of a plausible explanation for the efficacy variation patterns observed within the context of BCG-REVAC studies. Therefore, its quantitative conclusions shouldn't be extrapolated beyond this specific context. The analysis of new, hypothetical trials data similarly structured, conducted in other geographical areas, could certainly yield different results, both in terms of the values for the intrinsic vaccine efficacies and also in what regards the relative weights of masking and blocking mechanisms and their correlation with latitude. An additional limitation of BCG-REVAC data is due to the restriction of trials' endpoints to diseased and not diseased individuals, without measuring infection as a third relevant outcome. This limitation prevents us to address the relevant question of whether the vaccine is being blocked in its protective role against infection, or if, instead, blocking interferes more intensely with the vaccine's performance at reducing the progression rates from latency to active disease. As a general conclusion, more studies are needed to evaluate how general are the patterns found by BCG-revac trials, with the ultimate goal of assessing a positive explanation to the long lasting problem of BCG efficacy variation patterns. In this work, our aim is to provide a methodology useful for the analysis of such hypothetical studies.

The crucial implications of discriminating and quantifying masking and blocking effects for TB vaccine development are twofold. On one hand, understanding the range, and causes behind the variations of BCG efficacy is essential [31], as the efficacy of any novel vaccine will be measured against BCG. On the other hand, depending on where a new vaccine is applied and how old are the target populations, masking or blocking effects could affect new vaccines too.

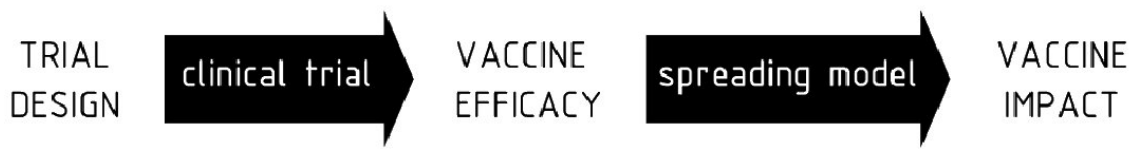

**Figure 6:** Scheme of the basis for evaluation of anti tuberculosis vaccines in absence of universally reliable protection correlates. First stage: design of vaccine efficacy determination clinical trials: the age of the cohorts must be elected taking into account that prior exposure to mycobacteria –either environmental, *M. tuberculosis* after exposure or even prior TST or also BCG– may corrupt the observed vaccine efficacies. Second stage: vaccine impact evaluations: bulk, short term and long term impact forecasts should be equally considered, as well as age-distributed impacts in terms of cases, infections and casualties prevented.

These issues affect different stages of the vaccine development pipeline, as sketched in figure 6. In the first place, during the process of vaccine evaluation in the context of clinical trials, studies of new tuberculosis vaccines should account for the possibility that prior infection may mask or (more likely) block their effects [5]. In this sense, and even if a new vaccine targeting TB in adolescents and adults rather than any other age group is expected to have the quickest impact on disease transmission and control, before we address the question of impact of novel vaccines, it is essential to know if the vaccine is more efficacious with respect to the current BCG. The most reliable way of knowing whether a new vaccine works better than BCG is by conducting an efficacy trial in a naive population without previous environmental sensitization (e.g., previous BCG vaccination, mycobacterial infection and/or TB contact) in order to avoid possible effects of masking or blocking [7,8,25]. It should be remarked that, beyond ES, as considered in this work, the very prior vaccination with BCG might trigger an additional blocking effect on further revaccination, specially when talking about live attenuated vaccines conceived as BCG's alternatives rather than as BCG boosters. This scenario would make clinical trials of these novel vaccines even more unreliable when performed on already vaccinated populations.

Furthermore, and once the efficacy estimation is complete, in order to produce any reliable vaccine impact and cost-effectiveness forecast, modelling scenarios contemplating ES deleterious effects on Tuberculosis vaccines are mandatory. The fact that, according to our analysis, blocking emerges as the driving effect behind BCG variability poses a potential pitfall to any vaccination strategy focused on individuals older than those analyzed here, very critically, most strategies conceived so far for booster vaccines. This is especially worth noticing because blocking, unlike masking, is not supposed to degrade the vaccine-induced protection obtained further during life by individuals immunized promptly after birth. Again, even if immunizing adolescents is thought to provide better impacts than vaccination strategies focused on younger age-segments, if such a novel vaccine is affected by blocking just as BCG is, then its impact will decrease in a way that, given the high blocking levels here identified, might even revert the comparison. As suggested by Helen McShane "we should optimize deployment of BCG to administration as close to birth as possible" [31], this should be the case for new priming live vaccines candidates based on BCG replacement strategies [6].

Taken all together, our results highlight the need for measuring ES effects on novel vaccines performance, as well as of diversifying vaccination strategies.

ACKNOLEWDGMENTS:

S.A. was supported by the FPI program of the Government of Aragón, Spain. JS was supported by the program of Postdoctoral Scholarships for Excellence of the Sainte-Justine UHC Foundation and by the Merit scholarship program for foreign students (PBEEE) of the Fonds de Recherche of Quebec, Nature et Tecnologies (FRQNT). This work has been partially supported by "Gobierno de Aragón/Fondo Social Europeo" and MINECO through Grant FIS2011-25167 to YM BIO2014-52580P, TBVAC2020 (643381) funded by the European Commission Horizon 2020 CM and DM; and the European FP7 grant NEWTBVAC 241745. Comunidad de Aragón (Spain) through FENOL to YM; and the EC Proactive project MULTIPLEX (contract no. 317532) to YM.

AUTHOR CONTRIBUTIONS:

S.A and J.S. designed the research and performed the simulations. S. A, J.S., D. M., C. M. and Y. M. analyzed and discussed the results. All authors wrote and approved the final version of the manuscript.